\documentclass[two column, nofootinbib, notitlepage]{revtex4-1}
\usepackage{epsfig,amssymb,amsmath,latexsym}
\usepackage{pifont}
\usepackage{hyperref}

\usepackage{graphicx}% Include figure files
\usepackage{dcolumn}% Align table columns on decimal point
\usepackage{bm}% bold math
\usepackage{epsfig}
\usepackage{color}

\usepackage{hyperref}

\hypersetup{
	colorlinks=true,
	linkcolor=red,
	citecolor=blue,
}

\usepackage{graphicx}
\usepackage{bm}

\begin{document}

\title{Testing Brans-Dicke Gravity with Screening by Scalar Gravitational Wave Memory}

\author{Kazuya Koyama$^1$}

\affiliation{$^1$Institute of Cosmology \& Gravitation, University of Portsmouth, Portsmouth, Hampshire, PO1 3FX, UK}
\date{\today}

\begin{abstract}
The Brans-Dicke theory of gravity is one of the oldest ideas to extend general relativity by introducing a non-minimal coupling between the scalar field and gravity. The Solar System tests put tight constraints on the theory. In order to evade these constraints, various screening mechanisms have been proposed. These screening mechanisms allow the scalar field to couple to matter as strongly as gravity in low density environments while suppressing it in the Solar System. The Vainshtein mechanism, which is found in various modified gravity models such as massive gravity, braneworld models and scalar tensor theories, suppresses the scalar field efficiently in the vicinity of a massive object. This makes it difficult to test these theories from gravitational wave observations. We point out that the recently found scalar gravitational wave memory effect, which is caused by a permanent change in spacetime geometry due to the collapse of a star to a back hole can be significantly enhanced in the Brans-Dicke theory of gravity with the Vainshtein mechanism. This provides a possibility to detect scalar gravitational waves by a network of three or more gravitational wave detectors. 

\end{abstract}

\maketitle

\section{Introduction}
The detection of gravitational waves has provided new possibilities to test the theory of gravity \cite{TheLIGOScientific:2016src, LIGOScientific:2019fpa}. Many modified gravity models introduce additional degrees of freedom in gravity and the detection of additional polarisations of gravitational waves will give a smoking gun for deviations from General Relativity (GR) \cite{Eardley:1973br}. In the scalar tensor theories, we expect the presence of the scalar gravitational waves \cite{Maggiore:1999wm, Nishizawa:2009bf, Callister:2017ocg}. 

However, scalar tensor theories are highly constrained by the Solar System measurements \cite{Will:2014kxa}. These constraints limit the strength of the coupling between the scalar field and matter. This makes it difficult to test these theories beyond the Solar System. Various screening mechanisms have been proposed to evade the Solar System constraints without suppressing the coupling \cite{Joyce:2014kja,Koyama:2015vza}. There are two main screening mechanisms. One is to suppress the scalar field gradient at the vicinity of an object. The representative example is the Vainshtein mechanism that can be found in various modified gravity models such as massive gravity, higher-dimensional braneworld model as well as scalar tensor theories \cite{Vainshtein:1972sx} (see \cite{Babichev:2013usa} for a review). The other example is the chameleon mechanism \cite{Khoury:2003aq, Khoury:2003rn} (see a review \cite{Burrage:2016bwy}). In this case, the scalar field mass changes depending on environments, suppressing the coupling between the scalar field and matter. The effectiveness of screening is determined by the spatial curvature of the object in the Vainshtein mechanism and the gravitational potential of the object in the chameleon mechanism. These screening mechanisms have distinct effects on structures in the Universe \cite{Falck:2015rsa}. Given the nature of these screening mechanisms to suppress deviations from GR in an environment with a large curvature/gravitational potential, we expect that deviations from GR will be suppressed further in a system such as binary black holes or neutron stars. This makes it difficult to test these models with gravitational wave observations \cite{Dar:2018dra}.  

Recently, a novel way to test scalar tensor theories was proposed using the gravitational wave memory \cite{Du:2016hww}. The gravitational wave memory is a permanent change in spacetime geometry \cite{1987Natur.327..123B}. In GR, this is caused by a burst event, which creates a jump in the transverse-traceless part of the space-time metric \cite{1974SvA....18...17Z, Christodoulou:1991cr}. In scalar tensor theories, the scalar mode of metric perturbations leads to a new scalar gravitational wave memory \cite{Du:2016hww}. Imagine a star that collapses to a black hole. Initially the scalar field is supported by the star. After the collapse, due to the no-hair theorem \cite{Hawking:1971vc}, the black hole does not support the scalar field. This causes a permanent change in the scalar component of metric perturbations outside the star, leading to the gravitational memory. It was shown that using a network of three or more detectors, it is possible to separate the scalar component of the gravitational waves \cite{Hayama:2012au} as different gravitational wave detectors on various locations have distinct responses to the different polarisations. Using a network of LIGO-Hanford, LIGO-Livingston \cite{LIGO}, Virgo \cite{VIRGO} and KAGRA \cite{KAGRA}, it is possible to detect the scalar gravitational memory from a collapse of a 10 $M_{\odot}$ star at the distance of 10 kpc can be detected in the Brans-Dicke theory by the second generation of gravitational wave detector network even imposing the Solar System constraint \cite{Du:2016hww}. The memory effect dominates the scalar stochastic gravitational wave background below 100Hz \cite{Du:2018txo}. 

The no-hair theorem applies to the scalar tensor theories \cite{Sotiriou:2011dz} as well as shift-symmetric galileon theories \cite{Hui:2012qt}, which accommodate the Vainshtein mechanism \cite{Nicolis:2008in}. A number of ways out exist for the black hole no-hair theorem (see \cite{Herdeiro:2015waa, Babichev:2016rlq} for reviews). For example, hairy black hole solutions can be found in galileon theories if the scalar field is allowed to acquire time-dependence \cite{Babichev:2013cya}. Hairy black holes can be found also in general relativity coupled with various matter fields as well as in a theory with a linear coupling of the scalar with the Gauss-Bonnet invariant \cite{Sotiriou:2013qea}. We will come back to this point in the discussion. 
  
In this paper, we show that the scalar memory effect provides a powerful way to test Brans-Dicke gravity with the Vainshtein mechanism. The Vainshtein mechanism is very difficult to test due to its very efficient suppression of the scalar force. The scalar memory effect will provide constraints that cannot be reached by the Solar System tests as well as astrophysical tests \cite{Sakstein:2017bws} and offer a possibility to discover scalar gravitational waves with high signal-to-noise ratio from a nearby gravitational collapse with a network of the second generation of gravitational wave detectors. We will also show that the scalar memory effect is suppressed in the chameleon mechanism.

\section{Brans-Dicke theory with Vainshtein mechanism} 
As a concrete example, we consider the model described by the following action \cite{Silva:2009km}
\begin{equation}
\int d^4x\sqrt{-g} \left[ \phi R - \frac{\omega}{\phi}(\partial \phi)^2 + f(\phi) \square\phi (\partial \phi)^2  + {\cal L}_m \right],
\label{action}
\end{equation}
where $\omega$ is the Brans-Dicke parameter, $(\partial \phi)^2 = \partial^\alpha \phi \partial_\alpha \phi$ and ${\cal L}_m$ is the matter lagrangian. The cubic interaction is the unique form of interactions at this order that keeps the field equation for $\phi$ of second-order \cite{Nicolis:2008in}. 

We expand the metric and the scalar field as $g_{\mu \nu} = \eta_{\mu \nu} + h_{\mu \nu}$ and $\phi = \phi_0 (1 + \varphi)$. By keeping non-linear terms in the second derivative of $\varphi$, which are relevant to the Vainshtein mechanism, the equations of motion become 
\begin{eqnarray}
\delta G_{\mu \nu} (h_{\mu \nu}) &=&  
(\partial_{\mu} \partial_{\nu} \varphi - \eta_{\mu \nu} 
\Box \varphi) + \frac{1}{2 \phi_0} \delta T_{\mu \nu}, \nonumber\\
(3 + 2 \omega) \Box \varphi &=&
-2 f(\phi_0) \phi_0^2 \Big[
(\partial_{\mu} \partial_{\nu} \varphi)(\partial^{\mu} \partial^{\nu} \varphi) - (\Box \varphi)^2 \Big] \nonumber\\
&&                                                                                                               + \frac{\delta T^{\mu}_{\mu}}{2 \phi_0},
\end{eqnarray}
where 
$\delta G_{\mu \nu} (h_{\mu \nu})$ is the linearised Einstein tensor and $\delta T_{\mu, \nu}$ is the linearised energy momentum tensor. 
Introducing $H_{\mu \nu} = h_{\mu \nu} - \eta_{\mu \nu}  \varphi$, we can diagonalise the equations for $H_{\mu \nu}$ and $\varphi$: $\delta G_{\mu \nu} (H_{\mu \nu}) = \delta T_{\mu \nu} / 2 \phi_0$. 

In the static limit, the scalar field equation becomes 
\begin{align}
\nabla^2 \varphi + r_c^2 
\Big[(\partial_{\i} \partial_{j} \varphi)(\partial^{i} \partial^{j} \varphi) -(\partial^2 \varphi)^2 \Big]
= - 8 \pi G \alpha^2  \rho,
\label{eom}
\end{align}
where we defined 
\begin{equation}
\phi_0 = (16 \pi G)^{-1}, \alpha = (2 \omega +3)^{-\frac{1}{2}},
r_c^2 = \frac{2 f(\phi_0) \phi_0^2}{3+ 2 \omega}.
\end{equation}
The parameter $\alpha$ controls the coupling between the scalar field and matter and $r_c$ controls the efficiency of the Vainshtein mechanism. 
The spherically symmetric solution is given by \cite{Schmidt:2009yj}
\begin{eqnarray}
\varphi (r) &&=   \frac{r_V^2}{4 r_c^2} g\left(\frac{r}{r_V} \right), \nonumber\\
 g(x) && = - \frac{x^2}{2} \left[-1 +  {}_2 F_1 \left(-\frac{2}{3}, - \frac{1}{2}, \frac{1}{3}, -x^{-3} \right) \right],
 \label{fx}
\end{eqnarray}
where
\begin{equation}
r_V = \left(16 \alpha^2 r_c^2 G M \right)^{1/3}
\end{equation}
is the Vainshtein radius and we imposed the condition that $\varphi(r) \to 0$ at $r \to \infty$. Fig.1 shows the profile of the scalar field. On larger scales $r > r_V$, the solution approaches the linear solution 
\begin{equation}
\varphi = \frac{2 \alpha^2 G M}{r}.
\label{linear}
\end{equation}
On the other hand, the scalar field is highly suppressed at $r \ll r_V$, realising the Vainshtein mechanism.  The important point is that the Vainshtein mechanism is not a screening mechanism in the standard sense. Normally the screening mechanism is a mechanism to reduce a charge. In the case of the Vainshtein mechanism, the scalar charge is given by $2 \alpha^2 M$ and this is not suppressed. The Vainshtein mechanism suppresses the scalar field gradient inside the Vainshtein radius due to the non-linear derivative interaction. Thanks to this feature, $\alpha$ can be O(1), i.e. the scalar force is as strong as gravity.   

\begin{figure}[h]
	\centering
	\includegraphics[width=8cm]{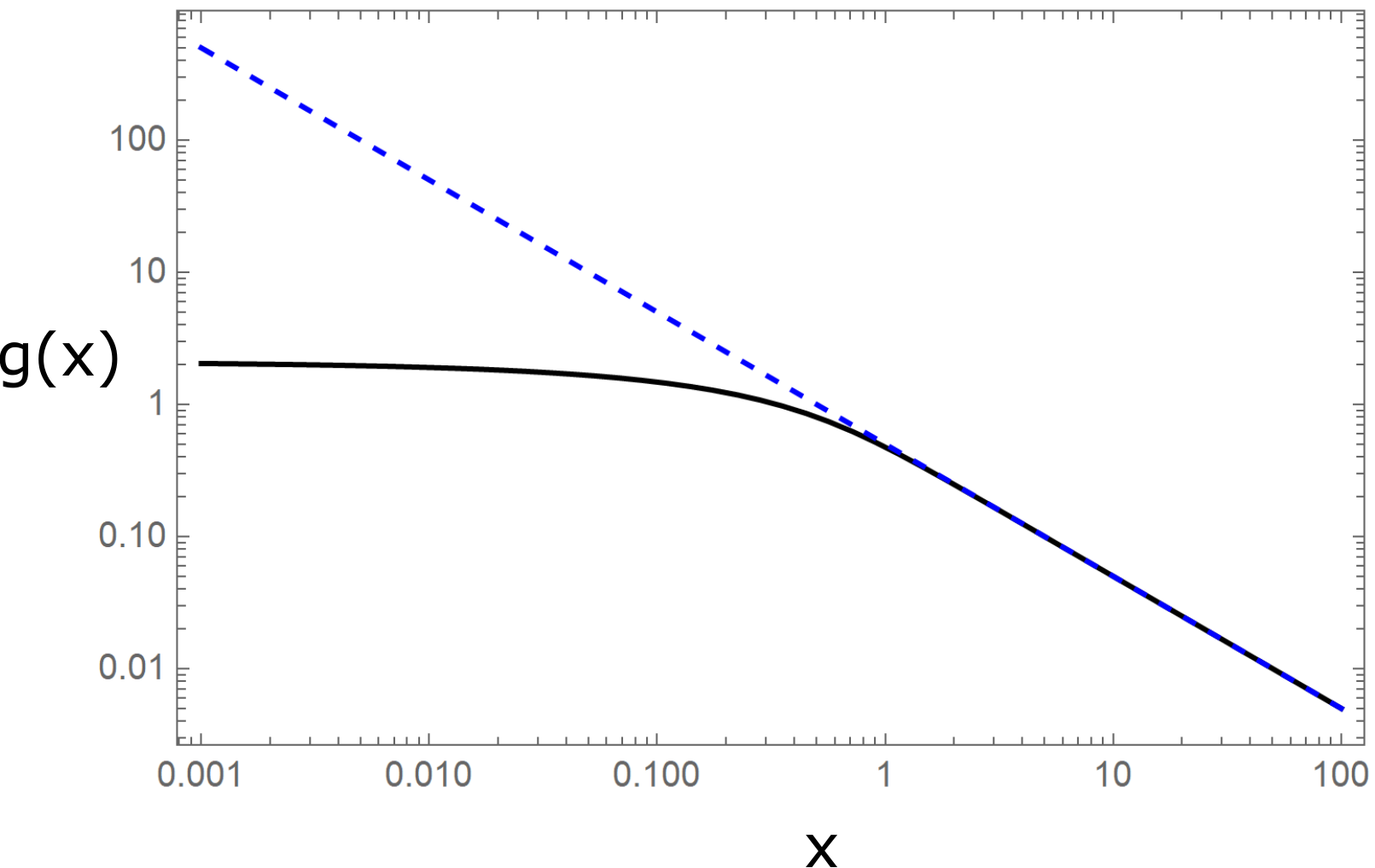}
	\label{Fig3}
	\caption{The function $g(x)$ in Eq.~(\ref{fx}). The dotted line shows the linear solution $2/x$ for $x \gg 1$. Below the Vainshtein radius $x<1$ the scalar field is suppressed due to the non-linear term.}
\end{figure}  

\section{Scalar gravitational wave memory}
Now, we consider the gravitational collapse of a star that forms a black hole. The no hair theorem of black holes states that black holes only support a trivial scalar field $\phi = \phi_0 =$ const. In fact, $\phi=$ const. is a solution without matter in our theory. Assuming that the no-hair theorem holds and the scalar hair is lost, the scalar field solution becomes $\phi = \phi_0$ from $\phi = \phi_0 + \varphi(r)$. This causes a permanent change in the scalar component of the metric perturbation 
\begin{equation}
\Delta h^S_{ij} = \varphi(r) e^{o}_{ij},
\label{change}
\end{equation}
where $e^{o}_{ij}$ is the scalar polarisation tensor. As long as the Solar System is located beyond the Vainshtein radius of the collapsing star, the scalar field solution is given by Eq.~(\ref{linear}). Fig.~2 shows the Vainshtein radius for a star with $M=1 M_{\odot}$ and $10 M_{\odot}$ with $\alpha=1$. For $r_c = 1000$Mpc, the Vainshtein radius for the Sun is $0.1$kpc and the Solar system constraint is well satisfied. On the other hand, for $M=10 M_{\odot}$, the Vainshtein radius is still $0.2$kpc and the linear solution can be used safely in the Solar System. In order for the Solar system to be inside the Vainshtein radius of the star with $M=10 M_{\odot}$ at the distance of $10$kpc from the Sun, $r_c$ needs to be $3.6 \times 10^5$ Mpc, which is well beyond the current horizon scale of the universe. Thus, for a reasonable choice of the parameters, we can approximate $\varphi(r)$ on Earth by the linear solution. 
Fig.~3 shows the change of the scalar component of the metric perturbation as a function of $\alpha$ assuming the linear solution. 

\begin{figure}[h]
	\centering
	\includegraphics[width=8cm]{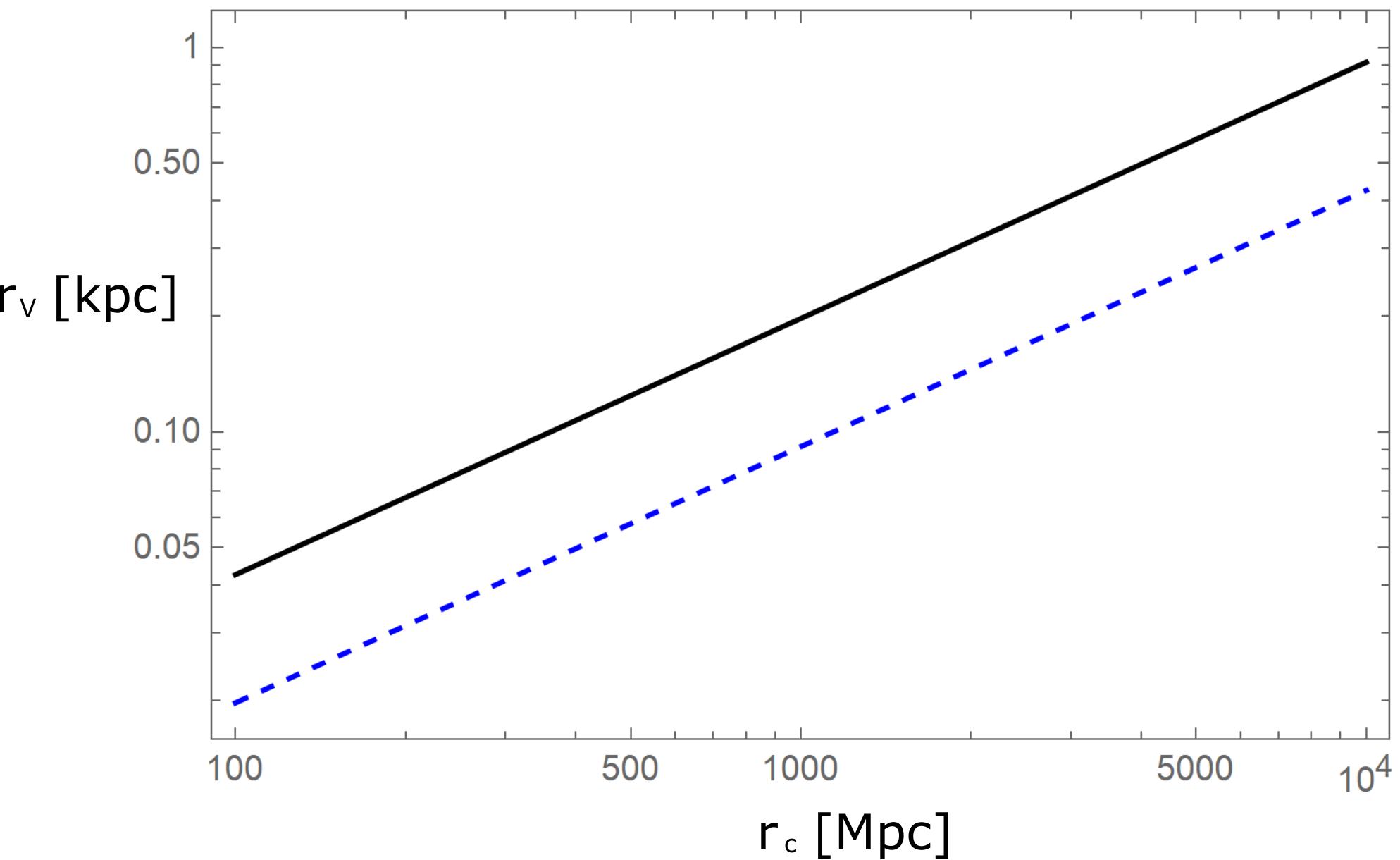}
	\label{Fig3}
	\caption{The Vainshtein radius $r_v$ in the unit of kpc as a function of $rc$ in the unit of Mpc for $\alpha=1$. The cosmological horizon scale today is $2998 h^{-1}$ Mpc where $h \sim 0.7$. The solid line is for a star with $M=10 M_{\odot}$ and the dashed line is for a solar mass star.}
\end{figure}

\begin{figure}[h]
	\centering
	\includegraphics[width=8cm]{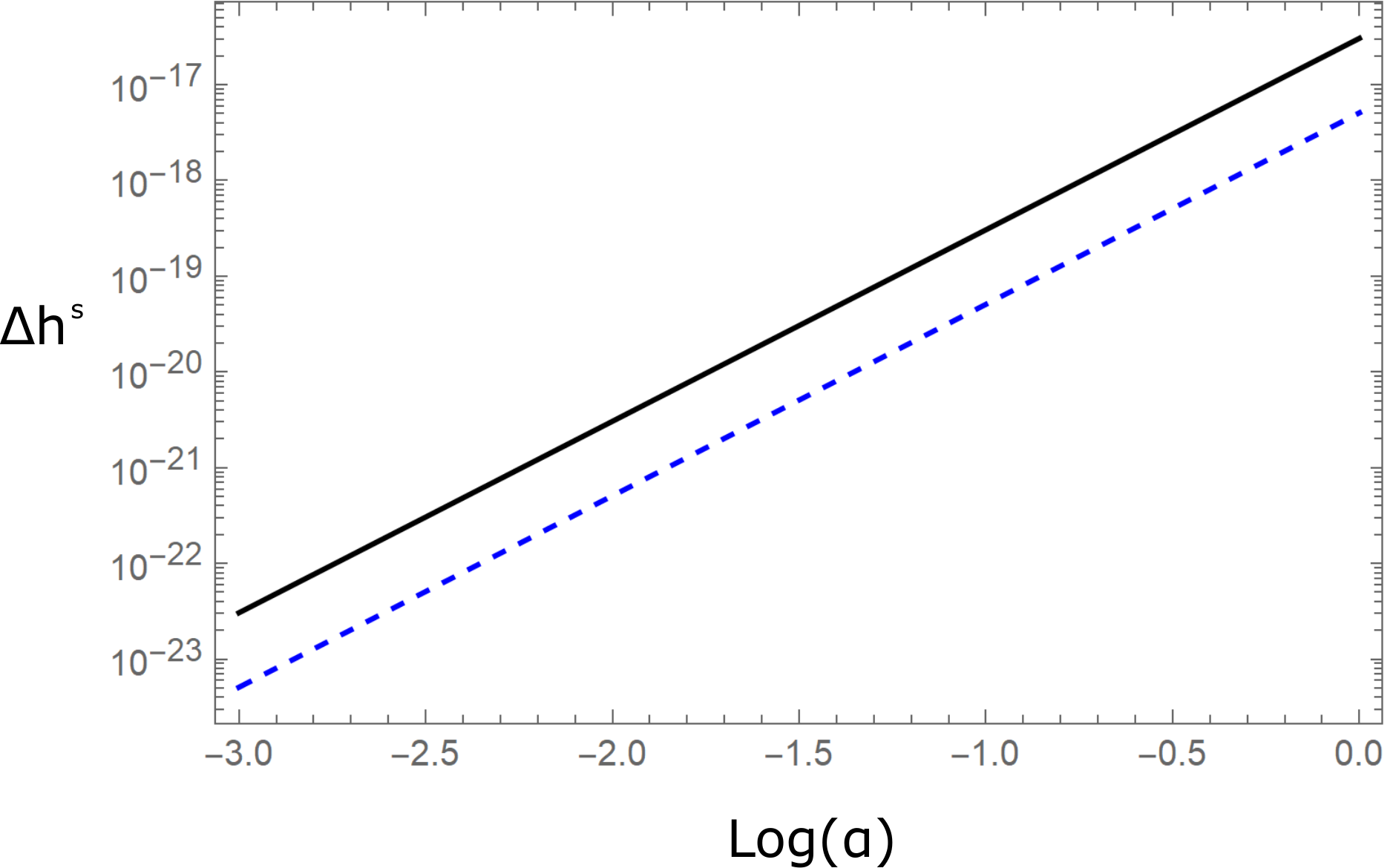}
	\label{Fig3}
	\caption{The change of the scalar component of metric perturbations given by Eq.~{change} as a function of $\alpha$. The solid line is for a star with $M=10 M_{\odot}$ at $10$kpc and the dashed line is for a star with $M=10 M_{\odot}$ at $60$kpc.}
\end{figure}  

The scalar gravitational wave memory has a frequency dependence of $h(f) \sim 1/f$ \cite{Du:2018txo}. The maximised Signal to Noise Ratio (SNR) for the detection of the scalar memory is given by \cite{Du:2016hww}
\begin{align}
\rho = {\cal F}_N^{1/2} \frac{2 \alpha^2 G M}{\pi r} \left[ 
\int^{f_c}_{0} df \frac{1}{f^2 S_n(f)}
  \right],
\end{align}	
where $S_n(f)$ is the noise spectral density that we assume all the detectors approximately have and $f_c$ is the cut-off frequency of memory determined by the time scale of the collapse. For a star with $M=10  M_{\odot}$ and $R=100 M_{\odot}$, this is estimated as $f_c = 500$ Hz. ${\cal F}_N$ is the N-detector effective angular pattern function and it is angular position dependent \cite{Hayama:2012au}. The angular averaged value for 3 (4) detectors is $F_3(4)=0.087 (0.240)$ and the peak value is $F_3(4)=0.511 (0.240)$. We use the noise spectral density given in Ref.~\cite{Sathyaprakash:2009xs} for the second and third generation detectors. Note that the detector configurations for the third generation detector considered in Ref.~\cite{Sathyaprakash:2009xs} are outdated but we used the same specifications that were used in \cite{Du:2016hww} to make a comparison easier. The scalar overlap reduction function for the Einstein telescope was computed in Ref.~\cite{Du:2018txo}. 

Fig.~4 shows the discovery curve defined as SNR=10 with $M=10 M_{\odot}$ as a function of the distance to the star. As shown in \cite{Du:2016hww}, the second generation detectors could discover the scalar gravitational memory if the star is 10 kpc away even in the Brans-Dicke gravity with no screening, for which the Solar System constraint imposes $\alpha < 10^{-2.45}$. If the distance is larger than 10 kpc, the third generation detectors would be required. On the other hand, the Vainshtein mechanism removes the limit on $\alpha$ from the Solar System constraint. The coupling can be as large as $\alpha \sim O(1)$. In this case, the second generation detectors would discover the scalar gravitational wave memory with $SNR>10^5$ as shown in Fig.~5.

\begin{figure}[h]
	\centering
	\includegraphics[width=8cm]{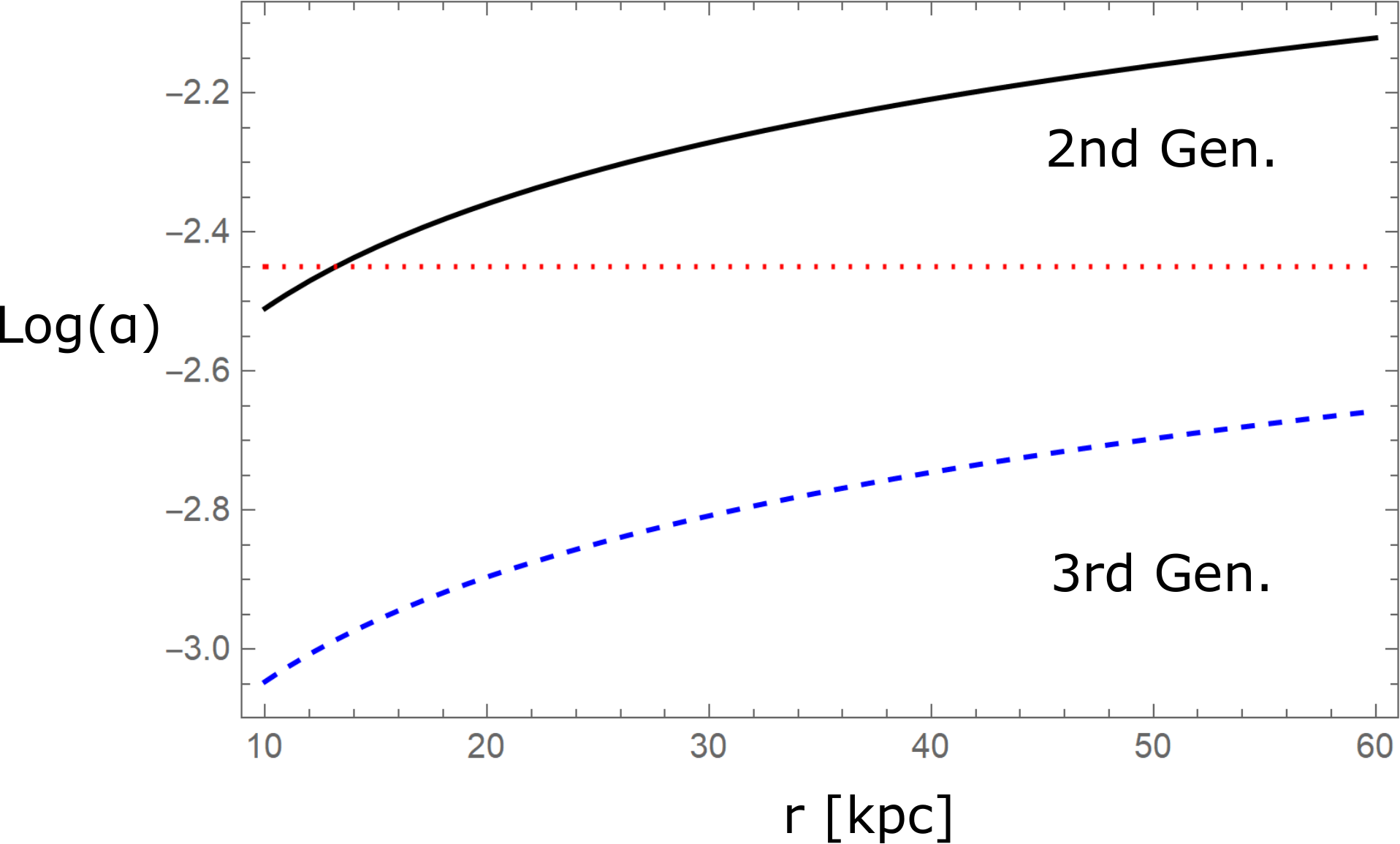}
	\label{Fig3}
	\caption{The discovery threshold for the scalar gravitational memory from a collapsing star with $M=10 M_{\odot}$ as a function of the distance from Earth in the unit of kpc. If $\alpha$ is above the curve, the signal will be detected with SNR=10. The solid (dashed) curve is for the 2nd (3rd) generation of detector network. The dotted line shows the Cassini bound (upper bound) in the Solar System as a reference. We emphasise that this bound does not apply to the theory with the Vainshtein mechanism and $\alpha$ can be $O(1)$.}
\end{figure}  

\begin{figure}[h]
	\centering
	\includegraphics[width=8cm]{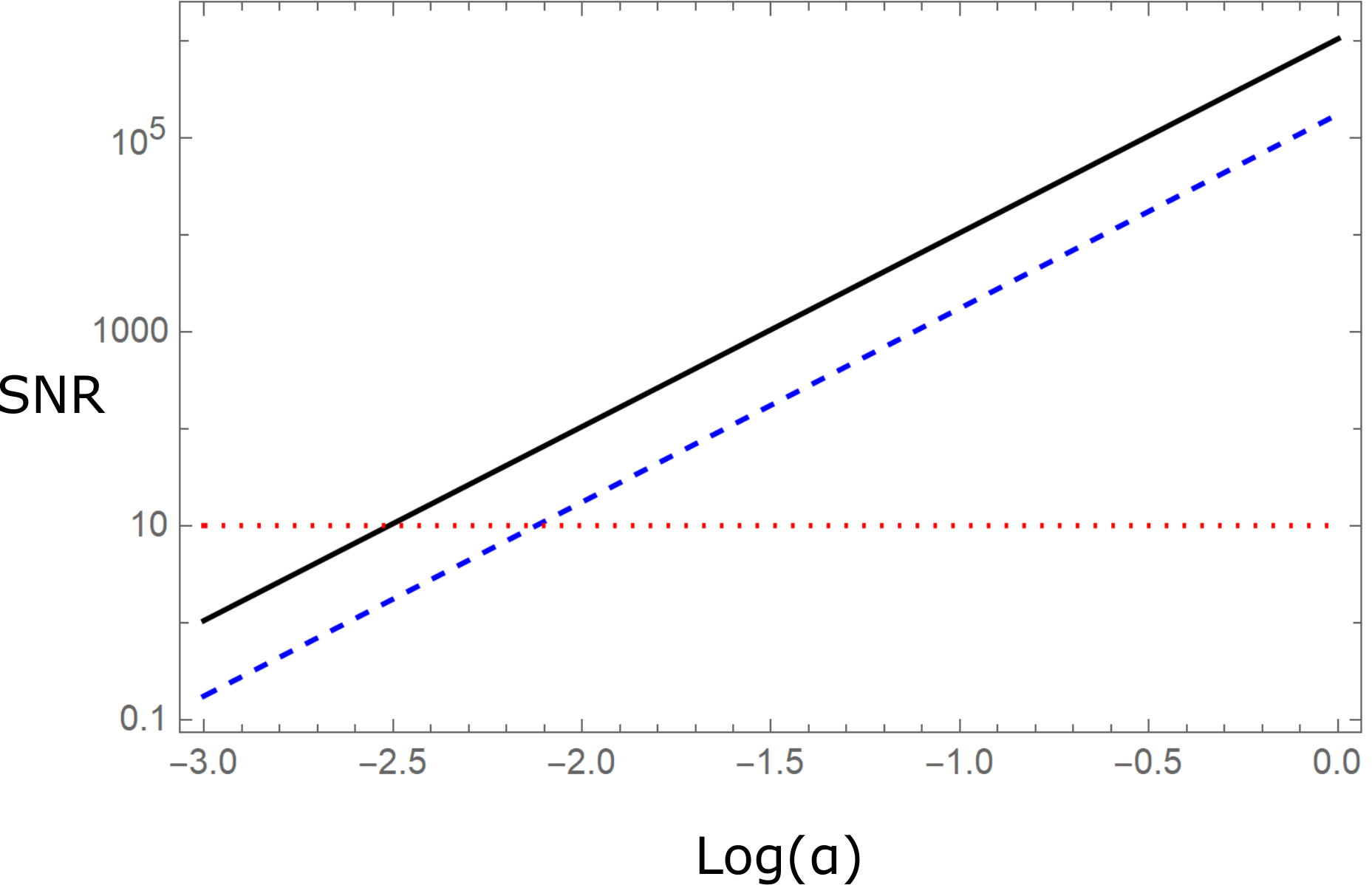}
	\label{Fig3}
	\caption{The SNR for the detection of the scalar gravitational wave memory in Brans-Dicke gravity with the Vainshtein mechanism from a collapsing star with $M=10 M_{\odot}$ at the distance of 10 kpc (solid line) and 60 kpc (dashed line) with the 2nd generation of detectors. This is valid as long as the Vainshtein radius of the star is less than 10 (60) kpc. The vertical dotted line shows SNR=10 used in Fig.~4.}
\end{figure}  

\section{Discussions}
We assumed that the Vainshtein mechanism of the Sun does not affect the scalar field profile generated by the distant star. This is a valid approximation thanks to the galileon symmetry of Eq.~(\ref{eom}) 
\cite{Hui:2009kc}. Since the equation contains only the second derivatives, it is possible to add a constant gradient. The scalar field gradient generated by the distant star can be approximated as constant at the vicinity of the Sun, and the solution can be written as 
\begin{equation}
\nabla \phi(r) = \nabla \phi_{\rm Sun}(r) + \nabla \phi_{\rm star}. 
\label{super}
\end{equation}
This is not the case if the Sun is inside the Vainshtein radius of the star \cite{Hiramatsu:2012xj}, but, in this case, the scalar field generated by the star is also suppressed. 

In this paper we used a weak field limit solution. The Vainshtein mechanism works also for relativistic stars \cite{Chagoya:2014fza, Ogawa:2019gjc} and the linear solution is recovered beyond the Vainshtein radius. As discussed in the introduction, black holes can have a scalar hair if the scalar field has a time dependence in galileon theories \cite{Babichev:2016rlq} although it is not clear whether hairy solutions can be formed as a result of the dynamical gravitational collapse or not. Since the existence of the non-trivial scalar field for a neutron star and the absence of it for a black hole are central to the scalar gravitational memory effect, it is important to derive neutron star solutions and prove the no-hair theorem for dynamically formed black holes in the theory described by Eq.~(\ref{action}). This is left for future work.     

It is difficult to test the Vainshtein mechanism with $r_c$ close to the horizon scale today. The lunar laser ranging and the observed precession of planets in the Solar System give constraints $r_c >$ a few hundreds Mpc \cite{Dvali:2002vf}. A better constraint is obtained by super massive black holes at the centre of a galaxy \cite{Sakstein:2017bws}. This test shares common ideas behind the scalar gravitational wave memory test based on the observation made in \cite{Hui:2009kc}. Due to the no-hair theorem of black holes, if part of a galaxy’s motion is due to an external scalar field sourced by a cluster where the galaxy is located, the supermassive black hole that lies at its centre does not feel this. On the other hand, for a galaxy located within a cluster, the scalar field sourced by the cluster behaves as a constant-gradient field as in Eq.~(\ref{super}) and the resident stars and dark matter of the galaxy respond to this cluster-sourced scalar field. Therefore the black hole lags behind as the galaxy moves. The absence of this lag put constraint on $r_c$. Although it is possible to get a better constraint than the Solar System test, for $\alpha = O(1)$, the constraint remains to be $r_c >$ several hundreds Mpc. The gravitational memory effect can prove the regime where $r_c$ is close to or even larger than the horizon scale. 

Finally we contrast the Vainshtein mechanism against another screening mechanism, the chameleon mechanism. In this model, the scalar charge is suppressed if the thin shell condition is satisfied, which is determined by the gravitational potential of the object. To satisfy the Solar System constraint, the Milky Way galaxy with the gravitational potential $10^{-5}$ needs to satisfy the thin shell condition. The collapsing star has a much larger gravitational potential and it is fully screened. Thus the scalar field sourced by the star is highly suppressed. Also the scalar field is screened on Earth. Unlike the Vainshtein mechanism, the superposition of scalar field gradients does not hold and if the scalar field is suppressed by Earth, then the external scalar field is also screened \cite{Hu:2009ua}. Thus we do not expect to see the scalar gravitational wave memory effect in theories with the chameleon mechanism.   

The scalar gravitational wave memory from a collapsing star has a potential to discover the Brans-Dicke theory of gravity that is indistinguishable from GR in the Solar System. In particular, the theory that utilise the Vainshtein mechanism will give a strong signal with SNR reaching $10^5$ even with the second generation of gravitational wave detector network. We need some luck with this test given the expected gravitational collapse rate of 2 events per 100 years within 60 kpc (note that this rate depends on star formation rates and other factors in the gravitational collapse \cite{New:2002ew}). However, if we detect the scalar gravitational waves together with the transverse traceless gravitational waves, this is a smoking gun of the theory beyond Einstein's theory of gravity. 

\section*{Acknowledgement}
KK thanks Jeremy Sakstein for useful discussions.  KK has received funding from the European Research Council
(ERC) under the European Union’s Horizon 2020 research and innovation programme (grant agreement No. 646702 ”CosTesGrav”) and the UK Science and Technologies Facilities Council grants ST/S000550/1.  

\bibliographystyle{unsrt}
\bibliography{reference}

\end{document}